\documentclass{kluwer}    % Specifies the document style.
\usepackage[dvips]{graphicx}

\newdisplay{guess}{Conjecture}
                               
\begin{document}                                                                                   
\begin{article}
\begin{opening}         
\title{Molecular Line Emission from Accretion Disks Around YSOs} 
\author{Itziar \surname{de Gregorio-Monsalvo}, Jos\'e F.\surname{G\'omez}}  
\institute{Laboratorio de Astrof\'{\i}sica Espacial y F\'{\i}sica Fundamental, INTA, Spain.}
\author{and Paola \surname{D'Alessio}}  
\institute{Instituto de Astronom\'{\i}a, UNAM, Mexico.}

\runningauthor{de Gregorio-Monsalvo, G\'omez and D'Alessio}
\runningtitle{Molecular Line Emission from Accretion Disks Around YSOs}

\date{June, 2003}

\begin{abstract} 
In this work we model the expected molecular emission from
protoplanetary disks, modifying different physical parameters, such as dust
grain size, mass accretion rate, viscosity, and disk radius, to obtain
observational signatures in these sources. Having in mind possible
future observations, we study correlations between physical parameters
and observational characteristics. Our aim is to determine the kind of
observations that will allow us to extract information about the
physical parameters of disks. We also present prospects for molecular
line observations of protoplanetary disks, using millimeter and
submillimeter interferometers (e.g., SMA or ALMA),
based on our results. 

\end{abstract}


\end{opening}           

\section{Introduction}  
                    % Produces section heading.  Lower-level
                    % sections are begun with similar 
                    % \subsection and \subsubsection commands.
%The formation of protoplanetary disks begins with a gravitational
%collapse in a dusty molecular cloud (Spitzer, 1982). The angular
%momentum of the gas is conserved during the contraction and the cloud
%gets an oblate shape. The material deposits in a disk perpendicular
%to angular momentum (Tereby et al. 1984), forming a protostar with a
%rotating disk (Shu et al. 1987). The gas continue falling onto the
%disk because of the centrifugal barrier produced by the angular
%momentum and the protostar acquire by accretion material from the
%disk (Walker et al. 1986). To be this falling possible it is
%necessary to free angular momentum and mechanic energy. This can be
%achieved by a disk with turbulent viscosity (Shakura and Sunyaev,
%1973) and the presence of winds (Rodriguez et al. 1982). 

The study of thermal molecular lines is fundamental to understand
structure and physical processes in protoplanetary disks. They are
emitted by the gaseous component of the disk and provide information
about kinematics, temperature and density of the cloud (Hartmann and
Kennyon 1987, Calvet et al. 1991, Najita et al. 1996). Today it is not
yet possible to resolve disks with this kind of lines with a good
signal-to-noise ratio, although it could be achieved with the next
generation of millimeter and submillimeter telescopes.

We expect protoplanetary disks to have a complex 3-D distribution of
the physical parameters that determine their molecular line emission
(e.g. density and temperature). These parameters, in their turn, will
depend on a variety of physical characteristics of the disk, the
central star, and their surrounding envelope.
Therefore, when making future observations of molecular lines in
protoplanetary disks, it may be difficult to derive physical
parameters from the observational characteristics, and in this derivation
will probably have to make use of a considerable amount of
assumptions.

In this work, we use the opposite approach: assuming a set of physical
parameters, we will try to predict which observational characteristics
yield more information about the former.

\section{Model and calculations}

%We have developed models of line emission in protoplanetary disks in
%order to estimate their future detectability with next generation of
%interferometers, i.e., SMA and ALMA. With our models, we try to
%reproduce the same observational characteristics that could be
%obtained in a real protoplanetary disk observation in the molecular
%transition chosen. 

In these initial calculations, we have assumed a young stellar object at a
distance of 140 pc (i.e., the distance to the Taurus cloud), 
with $60^{o}$ disk inclination angle, and stellar parameters typical
of T
Tauri stars: $M_{*}$= 0.5 M$_\odot$,
R$_{*}$=2R$_\odot$ and $T_{*}$= 4000 K. We have selected the (3-2) transition
of the C$^{17}$O isotope, because the envelopes of T Tauri stars are
likely to be optically thin for this line emission, given the low
relative abundance of this isotope.

\subsection{Model of protoplanetary disk structures}

To compute line emission using the transfer equation, we need a
detailed density and temperature model for every point within  the
disk. 
We have used D'Alessio et al. \citeyear{D'Al1998},
\citeyear{D'Al1999}, and  \citeyear{D'Al2001} models, which provide
these density  and temperature  
values in an self-consistent manner, making use of physical
equations. 

We obtained a network of models by varying different physical
parameters, like disk radius, viscosity parameter, accretion mass rate
and size of dust grains (Table \ref{table1}). 
Each family of input physical parameters,
determine a different density and temperature structure in the
disk. This structure is the input required in the solution of the
transfer equation.  

\subsection{Calculation of disk emission lines}

To solve the transfer equation, we divide the disk in cells, and the
emission is calculated along the line of vision, for each cell. 

After  obtaining the intensity of the emission at each velocity, it is
convolved with a particular beam, simulating a radio-telescope with a
0$\rlap.{''}$4 resolution. Afterward, the continuum emission  (originated  in
the dust)  is subtracted, to have the pure line emission (originated in
the gas). 

\begin{table} 
\caption[]{Physical parameter sets used in the models } 
\begin{tabular}{lllllll}                                       
\hline

\hline  
Disk radius (AU)                       & 50  & 100       & 150 & & &          \\
Maximum radius of dust grains ($\mu$m)     & 1   & 10        & 100  &10$^{3}$ &10$^{4}$ &10$^{5}$       \\
Mass accretion rate (M$_\odot$/year)          & 10$^{-9}$ &3 10$^{-8}$       &   10$^{-7}$  & & &     \\
$\alpha$ viscosity parameter                  & 0.001  & 0.005    & 0.01  &0.02 & 0.05       \\
\hline  
\hline
\end{tabular}
\label{table1}
\end{table}

\section{Results}

%From the resulting theoretical maps, we
%obtain several observational characteristics, trying to search for
%good tracers of physical parameters of disks. Such tracers could then
%be used in future observations in protoplanetary disks. To achieve
%this we performed a statistical treatment of our data, in particular,
%a  principal components and multiple regression analysis. 

\subsection{Emission maps}
\begin{figure}
\centerline{\includegraphics[height=10cm,angle=-90]{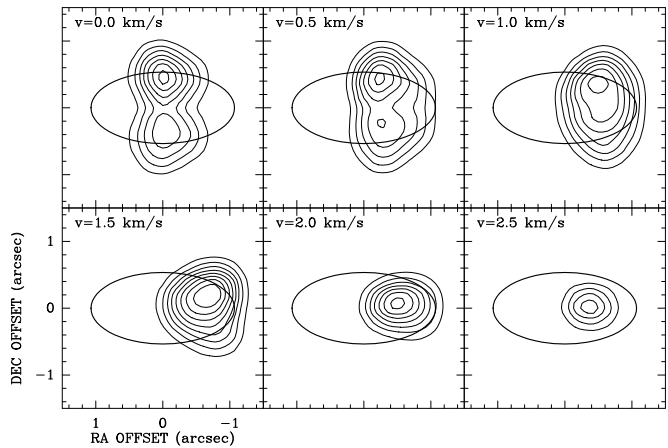}}
\caption{Emission maps at different velocities for a disk with radius
  = 150 AU, i = $60^{o}$, $\alpha$ = 0.01 and mass accretion rate =
  $10^{-7} M_{\odot}$/year. Maps have been convolved with a
  0$\rlap.{''}$4 beam. The lowest contour and the increment
  step are 20 mJy/beam. The ellipse traces the outer edge of the disk.} 
\label{fig1}
\end{figure}

The results of our calculations are maps of line emission, like the
ones shown in figure \ref{fig1}.

Qualitatively, we got the same characteristics as those obtained by
G\'omez and D'Alessio (2000). Our maps present a north-south asymmetry
(produced by the vertical structure in temperature of the disk) and  
an east-west asymmetry because of the
hyperfine structure of the C$^{17}$O molecule. The maximum emission peaks are
associated with the external parts of the disk at lower velocities,
because the increment of emitting area with radius, at those
velocities, is higher than the
decrement of the brightness temperature. Finally,  an intensity 
decrement in the central region is also seen. This is produced by the presence of
optically thick dust in the center, which yields a low contrast
between continuum and line.      

\subsection{Statistical Study}

From each map, we have measured the following observational
characteristics: intensity of the principal (northern) and secondary (southern) peak at each
velocity, 
distance from disk center to principal peaks, half power sizes 
of emission, and velocity at which the maximum intensity is present.
These represent a total of 43 different observational
parameters for each input model. In order to obtain the combinations
of 
such parameters that provide
information about initial physical characteristics, we have performed
a principal components analysis. 

\subsubsection{Principal Components}
This analysis reduces the total set of observational parameters (43 in
our case) to a smaller set of linearly independent parameters: the
principal components. These principal components are constructed as
linear combinations of the original parameters. We have found four significant principal
components, but the first two (PC1 and PC2) account for almost all the
variance within our sample of models (64\% and 18\%, respectively).

The physical parameters that define PC1 are (ordered by decreasing
relative weights) the velocity of the peak
emission, the half power sizes for principal peaks at intermediates
velocities and the distance from principal peaks to center.
The parameters that define PC2 are the half power sizes of principal
peaks at 1.5 km s$^{-1}$ velocity, the velocity of the peak emission
and the half power sizes of
secondary peaks at 0.0 and 1.5 km s$^{-1}$ velocities. 

\begin{figure}[H]
\tabcapfont
%\centerline{
\begin{tabular}{c@{\hspace{1pc}}c}
\includegraphics[width=2.3in]{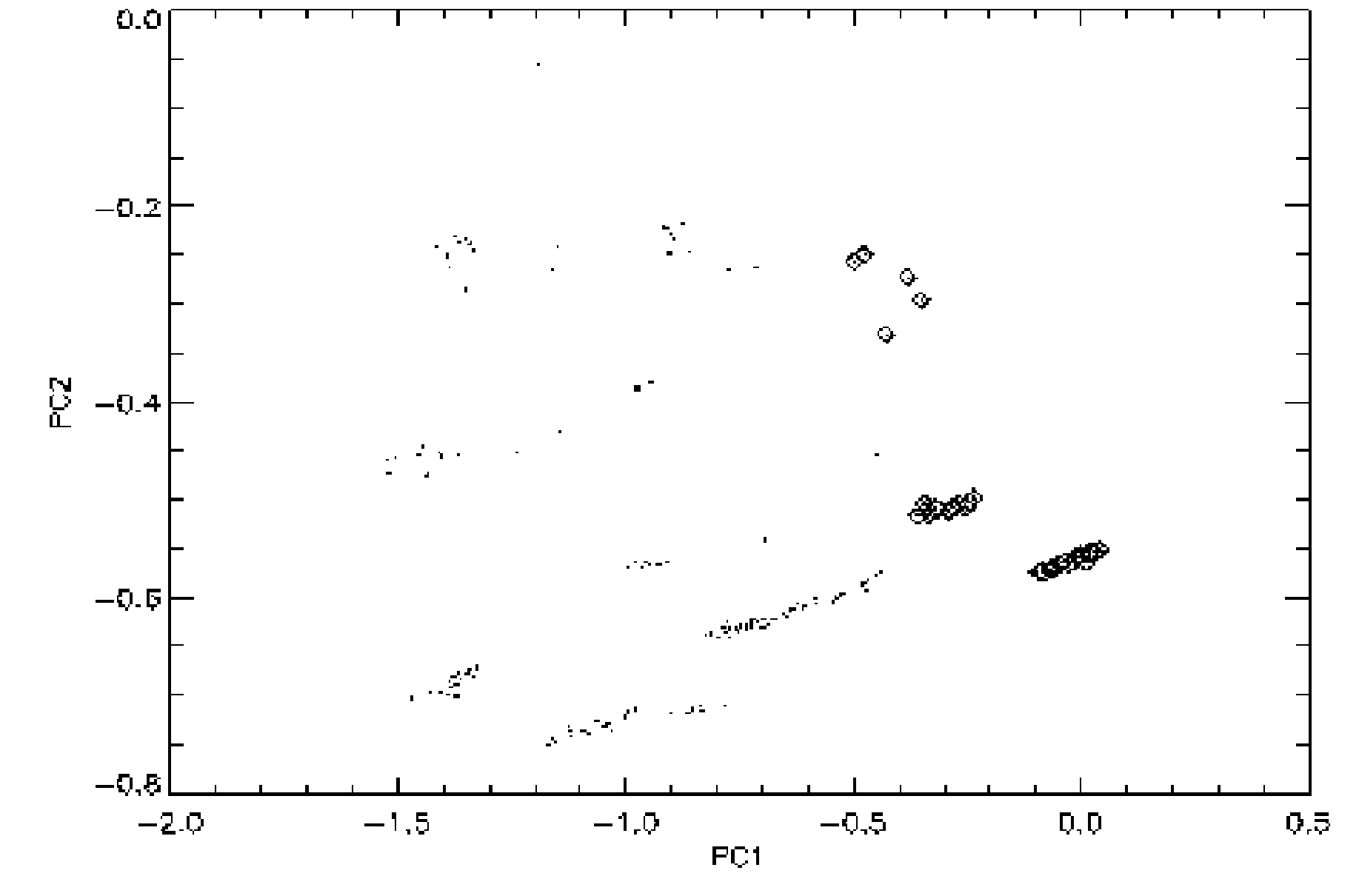} 
\includegraphics[width=2.3in]{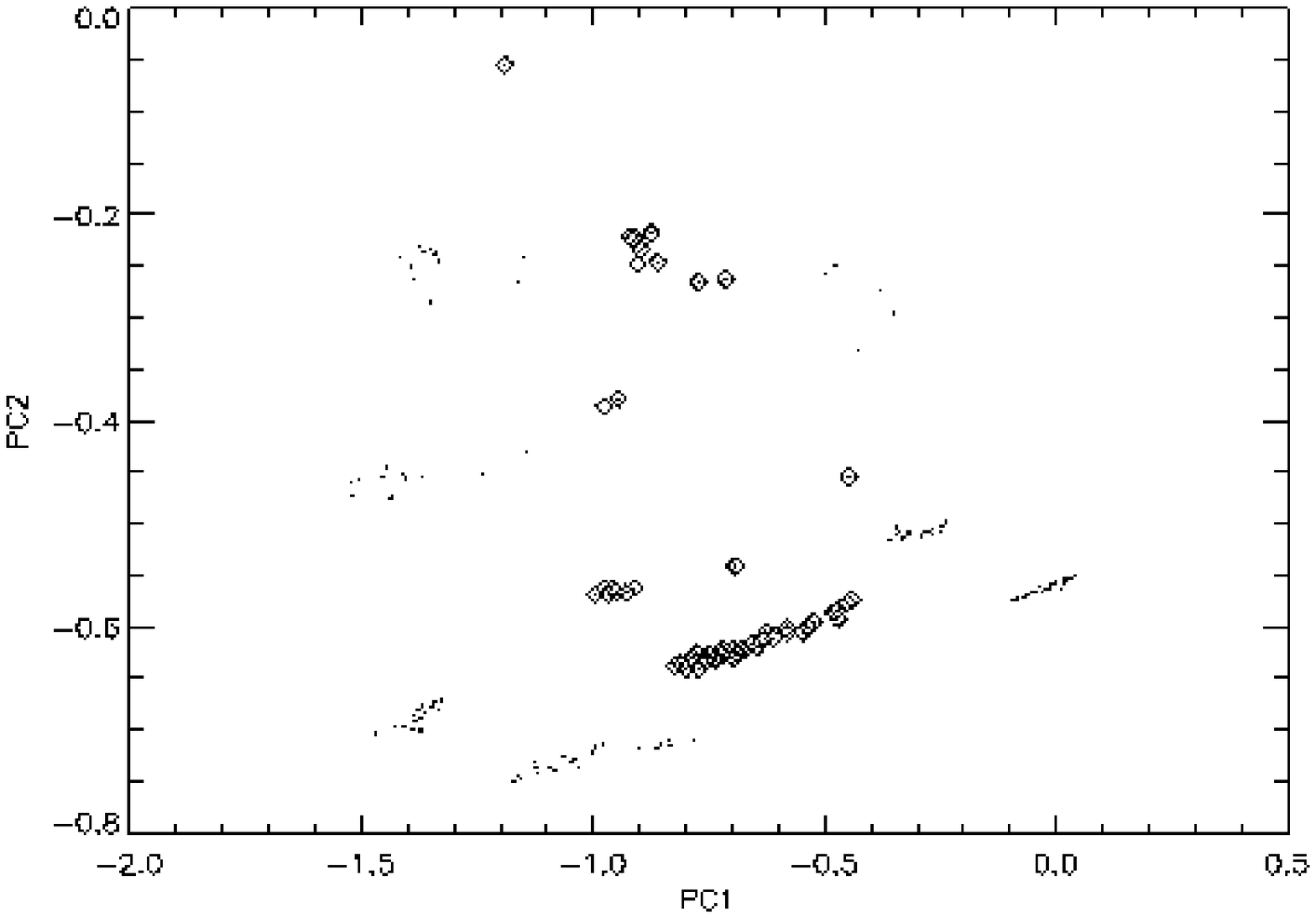}\\
\includegraphics[width=2.3in]{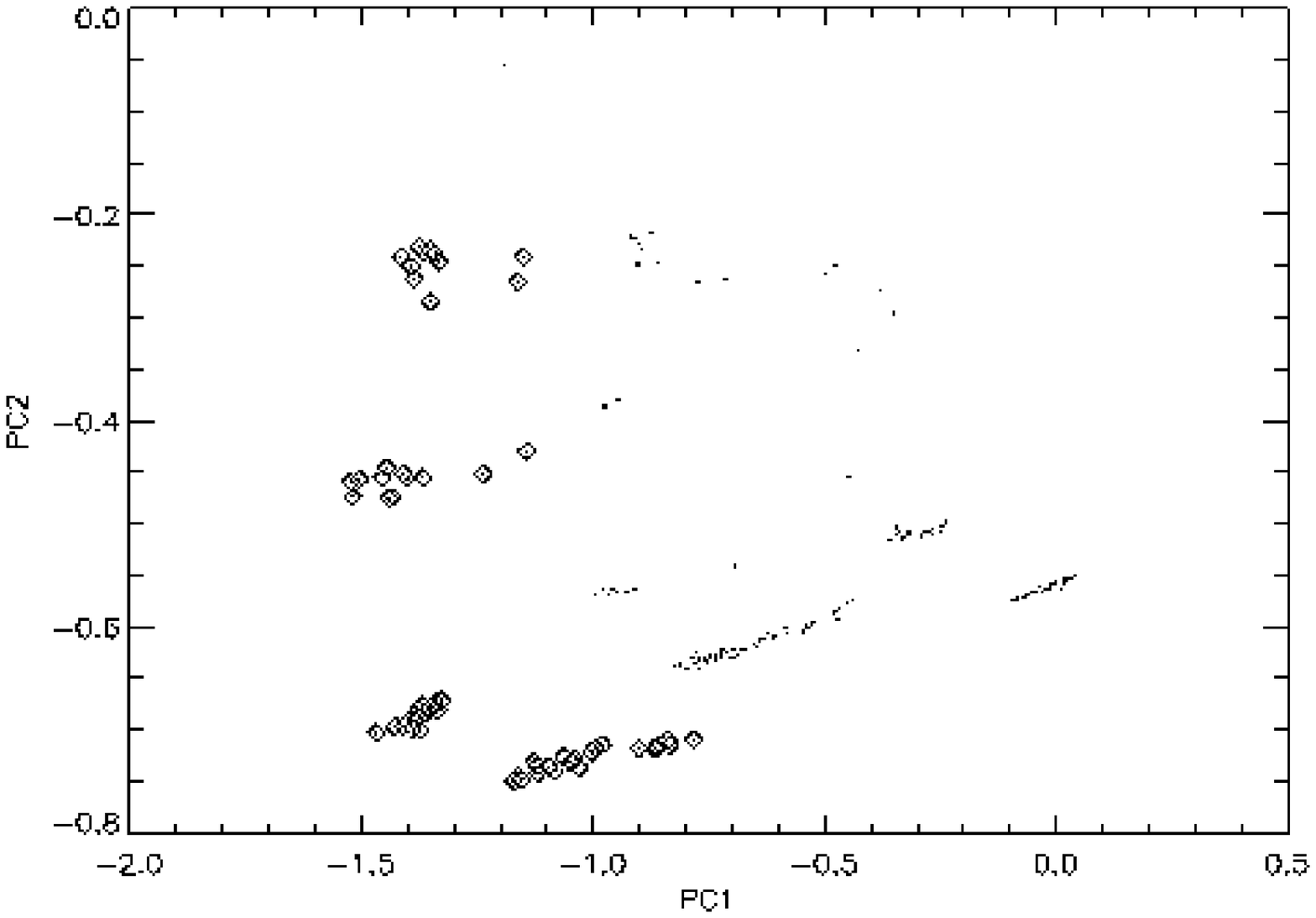}

\end{tabular}
\caption{PC1-PC2 diagrams. In all diagrams, dots represent the results
  for all calculated models. Squares represent disks with radius 50 AU
  (upper left panel), 100 AU (upper right) and 150 AU (bottom).}
\label{fig2}
\end{figure}

Analyzing PC1 and PC2 coefficients, we can see the variables that
provide most information about physical characteristics. The most
interesting result is shown when we represent the radius of the disks in a
PC1-PC2 diagram (see Figure \ref{fig2}), where it can be seen that the
first principal component (x axis) is good to discriminate among
different disk radii. Other trends relating principal components and
physical parameters are also present, but not so clearly.

\subsubsection{Multiple Correlation}

This study allows us to check whether we can quantitatively estimate each physical
parameter from  the set of observational variables. As an example, the correlation
analysis between principal components and both disk radius and mass
accretion rate gives:

$$R_d= -55.34 -0.94PC1 -0.18PC2 +0.16PC3 -0.08PC4$$
$$\dot{M}= 2\times10^{-7} -0.19PC1 +0.49PC2 -0.21PC3 -0.01PC4$$

The strongest correlation is obtained for radius (correlation
coefficient r=0.97), followed by mass accretion rate (r=0.57),
viscosity parameter (r=0.31) and maximum size of dust grains
(r=0.19). 

Looking at the relative coefficients in the equations above, this preliminary study suggests that PC1  provides a significant amount
of  information about radii, while PC2 provides some indication about
mass accretion rate (see their relative weights in the two correlation
expressions). On the other hand taking into account the physical parameters that give
rise to PC1 and PC2, we can conclude 
that a combination of the velocity of the peak
emission, the half power sizes for
principal peaks at intermediate velocities, and the distance from principal
peaks to disk center can give a good indication of disk radii. 
Furthermore, the half power sizes for secondary
peaks at 0.0 and 1.5 km s$^{-1}$, and of principal peaks at
1.5 km s$^{-1}$, can  provide some information about mass
accretion rates. 

We plan to extend this statistical study by enlarging the network of
input parameters, and calculating line emission for different molecules
and transitions.

\section{Detectability with SMA and ALMA}

With ALMA all the disks we have modeled can be detected at 5$\sigma$
with 1h integration time, and 1km s$^{-1}$ velocity resolution, in the
C$^{17}$O (3-2) transition. With SMA 
 all the disks modeled can be detected at the same conditions, except
 the one with 50 AU radius,   10$^{-9}$ M$_\odot$/year mass accretion
 rate and viscosity parameter 0.05. If integration time is raised to
 2h, all disks are detectable. These instruments will be useful
 tools to test  our predictions observationally.  

\begin{acknowledgements}

IdG and JFGacknowledge support from MCYT grant (FEDER funds) 
AYA2002-00376 (Spain). JFG is also
supported by MCYT grant AYA 2000-0912. 
IdG acknowledges the support of a Calvo Rod\'es
Fellowship from the Instituto Nacional de T\'ecnica Aeroespacial.
PD acknowledges grants from 
DGAPA-UNAM and CONACyT, M\'exico

\end{acknowledgements}

\end{article}
\end{document}